\documentclass[amsmath,amssymb,aps,twocolumn,prl,freq,floatfix]{revtex4-2}
\usepackage{graphicx}
\usepackage{dcolumn}
\usepackage[title]{appendix}
\usepackage{bm}
\usepackage{hyperref}
\usepackage{color}
\usepackage{subfigure}
\usepackage{flexisym}
\usepackage{physics}
\usepackage{dirtytalk}

\begin{document}

\title{Nitrogen vacancy center in diamond-based Faraday magnetometer}

\author{Reza Kashtiban}
\author{Gavin W. Morley}
\author{Mark E. Newton}
\author{A T M Anishur Rahman}
\altaffiliation{Department of Physics, University of Warwick, CV4 7AL, Coventry, UK}
\email{anishur.rahman@warwick.ac.uk}
%


\begin{abstract}
The nitrogen vacancy (NV) center in diamond is a versatile color center widely used for magnetometry, quantum computing, and quantum communications. In this article, using a single laser beam as a pump and probe, we measure the spin states of the NV center using the Faraday effect and use such measurements to develop a novel magnetic field sensor. Using the spin state-dependent effect on the left and right circularly polarized light, we probe and confirm the existence of spin-orbit coupling in the NV center at room temperature. The sensitivity of our magnetometer is $350~$nT/$\sqrt{Hz}$, limited by the background produced by the laser trapped inside the diamond. We argue that by using an optical cavity and a high-purity diamond, sensitivities in the femtotesla level can be achieved.
\end{abstract}

\maketitle







The ability to detect weak magnetic fields and their directions finds applications in fundamental physics including measuring the permanent electric dipole moments of electrons and neutrons \cite{BudkerRevModPhys2002,MurthyPRL1989,AbelPRA2020}, magnetic navigation \cite{Canciani2022,osti_1817974}, planetary exploration \cite{Korth2016,Sabaka2016}, magnetocardiography of our hearts \cite{BrisindaD.2023Cmtu} and magnetoencephalography of our brains \cite{Hämäläinen1993,BROOKES2022621}. One of the most sensitive vector magnetometers is an optically pumped magnetometer (OPM) \cite{DangAPL2010,KominisNat2003,Fabricant_2023} with a sensitivity of $160~$attotesla \cite{DangAPL2010}. Such a sensor uses the Faraday effect-based detection scheme, operates in a magnetically shielded environment, and can resolve magnetic fields with a centimeter-scale resolution (important in biomedical imaging \cite{BrisindaD.2023Cmtu,Hämäläinen1993,BROOKES2022621}). Scalar OPMs can operate in ambient conditions but suffer from heading errors \cite{Fabricant_2023, Canciani2022}. Superconducting quantum interference devices (SQUIDs) are another class of highly sensitive magnetometers that provide femtotesla-level sensitivity \cite{Buchner2018}, but they require cryogenic cooling.



\begin{figure}
    \centering
    \includegraphics[width=8.50cm]{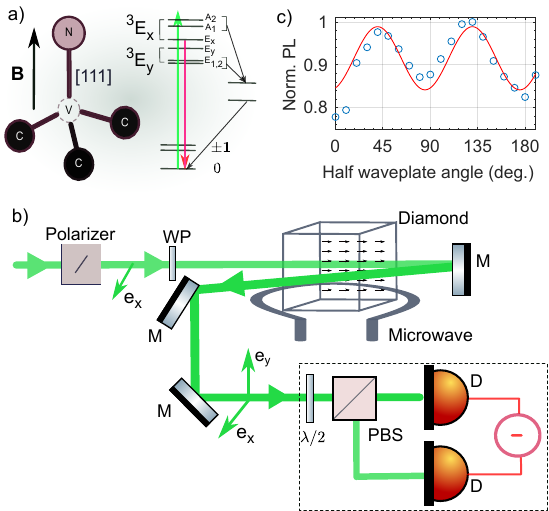}
    \caption{a) The nitrogen-vacancy (NV) center in diamond and its energy level diagram at low temperature. Here, an NV center oriented along the $[111]$ direction has been shown. A dc magnetic field $B$ is applied along the $[111]$ crystallographic direction for lifting the degeneracy of the NV center's ground state manifold. b) A schematic of our experimental setup. A green laser ($532~$ nm) is used for exciting the NV centers from their ground state manifold ($^3A_2$) to the excited state manifold ($^3E$). Before interacting with the NV centers, the green laser passes through a linear polarizer and a half/quarter wave plate (WP). The axis of the polarizer is parallel to the $x~$axis of the laboratory frame. A mirror located after the diamond sends back the green laser through the diamond for a second time. A pick-off mirror is used to collect the retroreflected laser beam. Other optical components of the experiment are: $M$ - mirror, $D$ - photodiode, PBS - polarizing beam splitter, and $\lambda/2$ - half waveplate. Components within the dotted box make the balanced polarimeter. c) Photoluminescence as a function of the angle of the half-wave plate located before the diamond.}
    \label{fig1}
\end{figure}

The nitrogen vacancy center in diamond is a versatile atom-like system \cite{Doherty2013}, used for magnetometry \cite{BarryRevModPhys2020,WolfPRX2015,AcostaAPL2010,Chatzidrosos2017,GrahmanPRR2023}, and has been proposed for quantum computing \cite{Pezzagna2021} and  quantum gravity experiments \cite{BosePRL2017}. The NV center-based magnetometers \cite{WolfPRX2015,AcostaAPL2010,Chatzidrosos2017,GrahmanPRR2023} operate in ambient conditions, are immune to heading errors, and can spatially resolve magnetic fields with a millimeter-scale resolution \cite{GrahmanPRR2023,Wolf2015}. However, the sensitivity of current NV magnetic field sensors \cite{WolfPRX2015,AcostaAPL2010,Chatzidrosos2017,GrahmanPRR2023} is on the order of a picotesla and not suitable for demanding applications such as magnetoencephalography. One of the main limitations of NV magnetometers is their detection scheme, which uses the NV spin-state-dependent photoluminescence (PL), a spin state destructive measurement system. Collecting PL emitted in all directions by NV centers is challenging. Detection schemes that can measure the spin states of NV centers non-destructively, as well as repeatedly and without relying on photoluminescence, would benefit magnetometry, quantum computing, and quantum gravity experiments.



In this article, we measure the spin states of NV centers using the Faraday effect, a non-destructive measurement scheme \cite{BuckleySci2010,ShahPRL2010}, and using such measurements develop a magnetometer with the potential to reach femtotesla sensitivity. We investigate different optical transition dipole selection rules using linearly and circularly polarized light and confirm the existence of spin-orbit coupling in NV centers at room temperature.

The nitrogen-vacancy center in diamond consists of a substitutional nitrogen atom and an adjacent vacancy in the diamond lattice. Figure \ref{fig1}a shows an NV center oriented along the $[111]$ crystallographic axis. The diamond used in our experiment is a single crystal $\{111\}$ cut $2\times2\times2~$mm$^3$ Type 1b high-pressure high temperature (HPHT) single-crystal synthetic diamond, and contains $0.220~$ppm NV centers and $54~$ppm substitutional nitrogen. For magnetometry, we apply a homogeneous dc magnetic field ($B$) parallel to the $[111]$ axis of the diamond. The ground state manifold ($^3A_2$) of an NV center is an orbital-singlet spin-triplet ($S=1,~m_s=0,~\pm 1$) state with zero orbital angular momentum, while the excited-state manifold is an orbital doublet and has orbital angular momenta of $\pm 1$ \cite{ToganE2010,Maze_2011}. As we use an ensemble of NV centers, strain is inevitable \cite{Maze_2011,Doherty2013,FuSantoriPRL2009,MansonPRB2006,BatalovPRL2009}, which splits the excited state into two branches: $^3E_x ~(A_1,~A_2,~E_x$) and $^3E_y~(E_1, E_2,~E_y)$.  At relatively low strain the $A_1,~A_2,~E_1,~E_2$ and $E_y$ states are spin-orbital entangled states e.g., $A_2=|E_+\rangle \otimes |-1\rangle +|E_-\rangle \otimes |+1\rangle$ \cite{Maze_2011,ToganE2010}, where $E_\pm$ is the excited state orbital angular momentum and $|\pm 1\rangle$ represent spin states for $m_s=\pm 1$. $E_{x,y}$ states only couple to the $m_s=0$ state of the ground state manifold by absorbing linearly polarized light, while the other four states 
 can be excited from the $m_s=\pm 1$ states via circularly polarized light.

A schematic of our experimental setup is shown in Fig. \ref{fig1}b. The diamond is oriented such that the $[111]$ crystallographic direction is along the $z-$axis. A linearly polarized green laser ($532~$nm) propagating along the $z~$axis non-resonantly excites the nitrogen-vacancy center from the ground state manifold to the excited state manifold and eventually spin polarizes the NV centers to the $m_s=0$ state \cite{Doherty2013}. Due to the ensemble nature of our experiment, we excite NV centers of all orientations along the path of our excitation beam. As well as using the green laser as an NV center excitation source, we use it as a probe to measure the spin states of NV centers via the Faraday effect (FE). To excite more NV centers and enhance the Faraday effect, the green laser is retroreflected and picked up using a pickoff mirror. The double-passed green laser is then inputted to a balanced polarimeter composed of a half-wave plate, a polarizing beam splitter, and a balanced photodetector consisting of two identical photodiodes \cite{ShahPRA2009,BudkerRevModPhys2002}. The retroreflected green laser is also used for detecting absorption detected magnetic resonance (ADMR) \cite{AhmadiPRB2017}. In this case, the retroreflected green laser is directly inputted to one of the photodiodes of the balanced detector, while the other photodiode receives a fraction of the green laser that has not interacted with the diamond. Additionally, photoluminescence emitted by NV centers is detected using a detector placed along the $x~$axis.




In our first experiment, we measure PL as we rotate the plane of polarization of the green laser using a half-wave plate placed immediately after the polarizer and no microwaves are used. The axis of the polarizer is parallel to the $x~$axis, which we use as a reference. Two orthogonal electric dipoles associated with the NV centers along the $[111]$ axis are in the $x-y$ plane while the remaining six dipoles belonging to the other NV centers are in planes that make an angle $\approx 71^o$ with the $x-y$ plane \cite{Reuschel2022}. Figure \ref{fig1}c shows the results of our measurements \cite{EpsteinNat2005,FuSantoriPRL2009}. Due to the non-resonant and ensemble nature of our experiment, identifying the orientations of $^3E_x$ and $^3E_y$ unambiguously is challenging. Nevertheless, taking into account the dipole radiation pattern, we fit a function of the form $A-B\cos{2\alpha}-C \cos{(2\alpha-4\phi)}$ and retrieve $\phi$ - the angle between the $x~$axis and $^3E_x$. The angle between the $x~$axis and the laser polarization is denoted by $\alpha$. From fit we find $\phi\approx 72^o$ indicating that $^3E_x$ and $^3E_y$ are located at $\phi=72^o$ and $\phi=-18^o$, respectively. Note that because of the position of our PL detector and the small angle over which we collect PL, the total photoluminescence is independent of the orientations of other dipoles (see supplementary information).





\begin{figure}
    \centering
    \includegraphics[width=8.5cm]{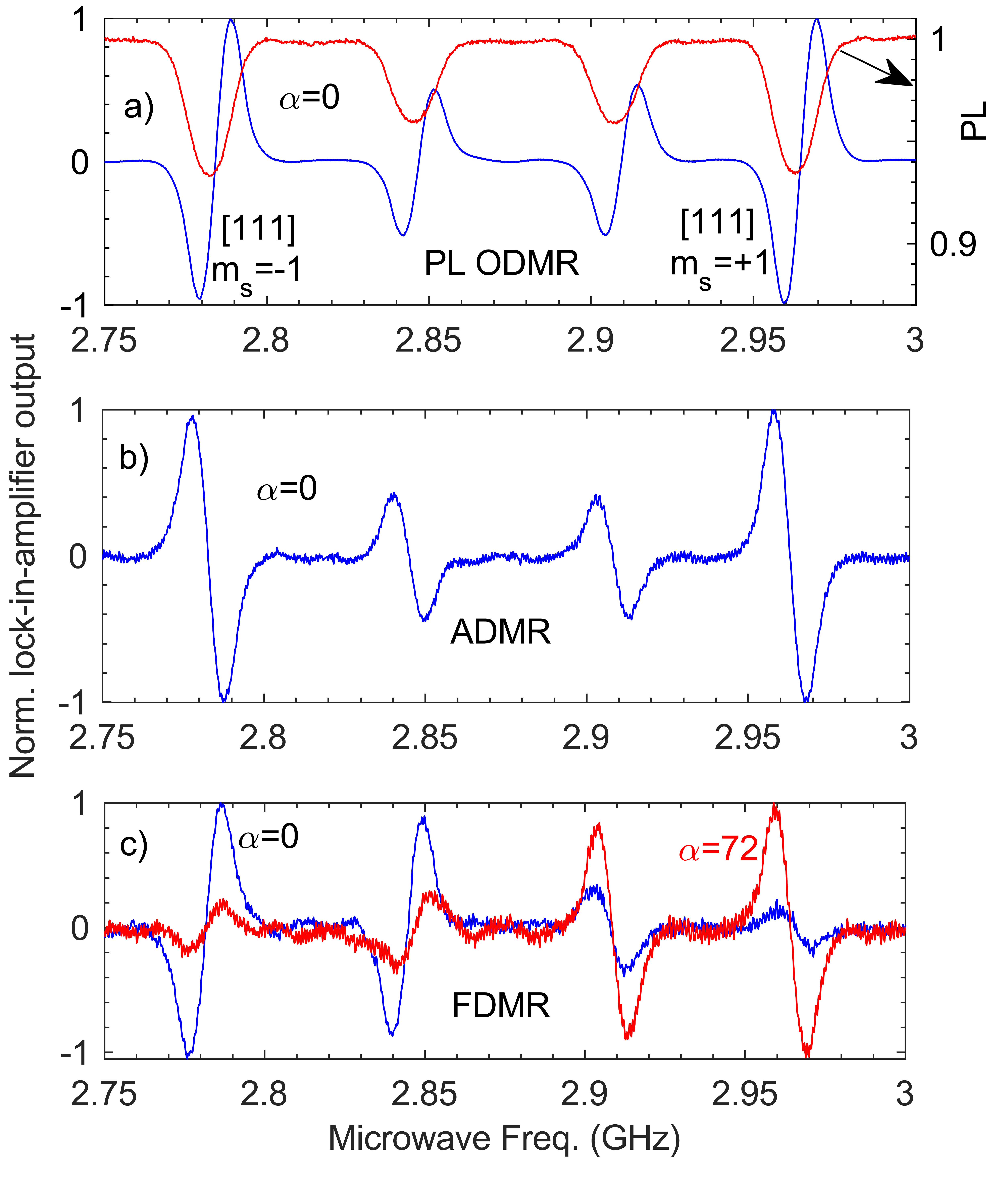}
   \caption{\textbf{Magnetic resonances} a) Photoluminescence-based optically detected magnetic resonance (ODMR) when the angle between the linear polarizer and the half-wave plate after it is zero (($\alpha=0$). The blue data uses a lock-in amplifier while the red data do not. b) The green laser absorption detected magnetic resonance (ADMR) c) The Faraday detected magnetic resonance (FDMR) for $\alpha=0 ~\& ~72^o$.}
    \label{fig2}
\end{figure}





Figure \ref{fig2}a shows the raw photoluminescence (red) and the same when detected using a lock-in amplifier (blue) as a function of microwave frequency. The polarization of the laser is parallel to the $x~$axis ( $\alpha=0$). Due to $B\parallel [111]$, the ODMRs associated with the NV centers along the $[111]$ crystallographic direction appear at $(D+\gamma m_sB)/2\pi$ while the ODMRs of the remaining NV centers cluster at $(D+\gamma m_sB \cos\theta)/2\pi$ with $\theta\approx 109^o$. The ODMRs at $(D+m_s\gamma B)/2\pi$ are stronger than others due to the orthogonality of the microwave magnetic field with these centers, allowing efficient excitation. Fig. \ref{fig2}b shows ADMR \cite{AhmadiPRB2017}. The polarities of all ADMR peaks are identical but are opposite to those due to PL.



Figure \ref{fig2}c shows the output of the lock-in amplifier when the retroreflected green laser and the balanced polarimeter are in use. Four peaks are visible, which we call the Faraday detected magnetic resonance (FDMR). The two peaks belonging to the $m_s=0\rightarrow -1$ transitions at lower microwave frequencies are stronger than those due to the $m_s=0\rightarrow +1$ transitions for $\alpha=0$ (blue). For comparison, we also show data when the polarization of the laser is parallel to the $^3E_x$ dipole ($\alpha=72^o$) (red). The peaks belonging to the $m_s=0\rightarrow +1$ transitions are now dominant. Most importantly, in both cases, the polarities of the $m_s=0\rightarrow -1$ transitions and the $m_s=0\rightarrow +1$ transitions are opposite. This contrasts with ADMR and ODMR, which have peaks of the same polarities. Linearly polarized light is composed of left ($\hat{\sigma}_+$) and right ($\hat{\sigma}_-$) circularly polarized light i.e., $\hat{x}=(\hat{\sigma}_++\hat{\sigma}_-)/\sqrt{2}$ and $\hat{y}=-i(\hat{\sigma}_+-\hat{\sigma}_-)/\sqrt{2}$. When $\alpha=72^o$, we only excite the $^3E_x$ dipole due to the optical selection rules \cite{Maze_2011}. NV centers in the $m_s=-1$ state add a phase to the $\hat{\sigma}_+$ component of the linearly polarized light, keeping the $\hat{\sigma}_-$ component intact \cite{BuckleySci2010,HuRarity2008,Atatur2007}. The spin state-dependent phase rotates the plane of polarization of linearly polarized light in the opposite direction, giving rise to FDMR peaks of opposite polarities. We note that in our Faraday measurements, the $^3E_y$ dipole appears at $\alpha=0$ instead of $\alpha=-18^o$ predicted by the fluorescence measurement (Fig. \ref{fig1}c). This might be due to the strain-induced reorientation of the dipole.

\begin{figure}
    \centering
    \includegraphics[width=8.5cm]{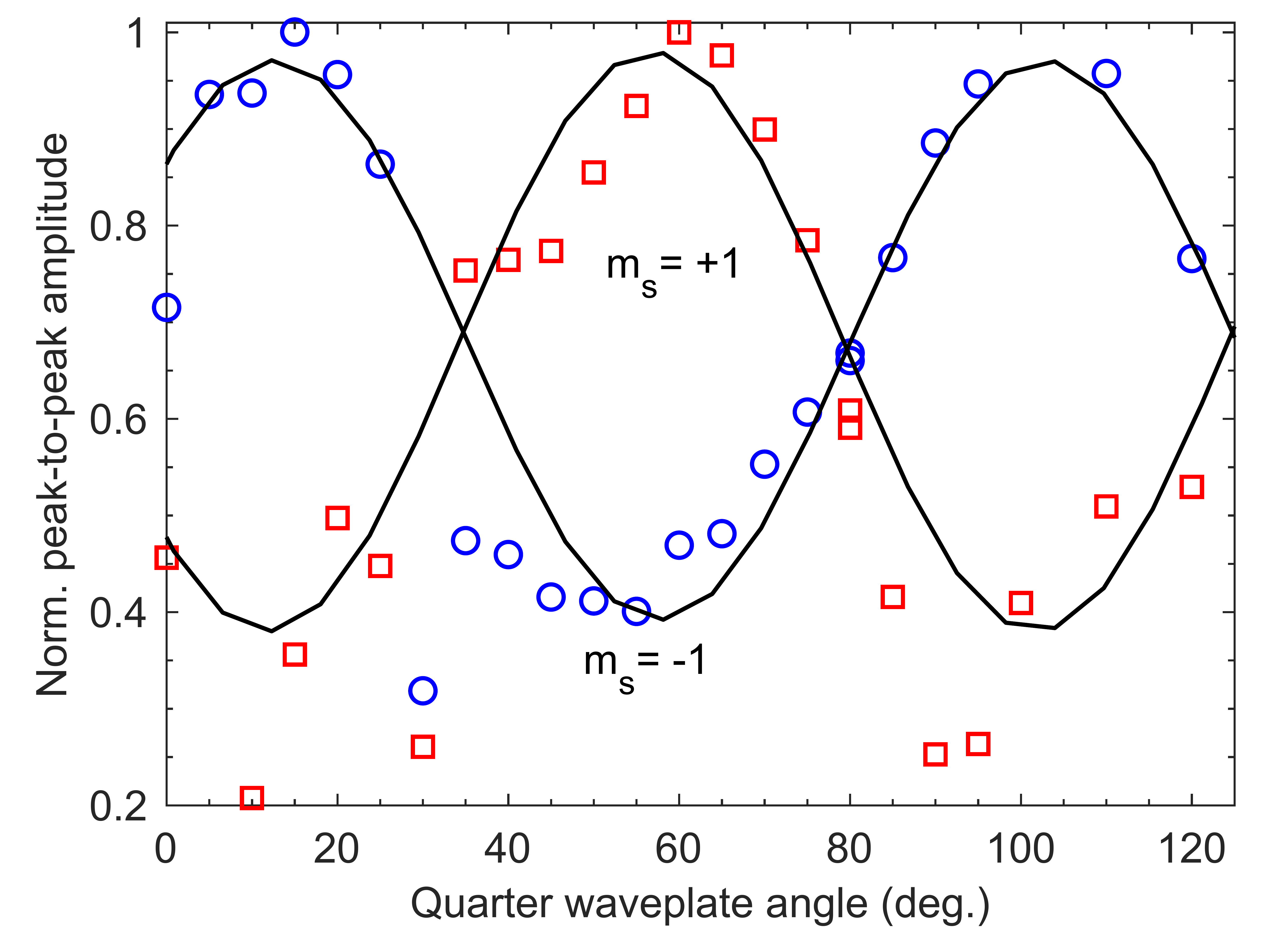}
    \caption{The peak-to-peak amplitude of FDMR as a function of quarter-waveplate angle $\alpha$ when the $m_s=0\rightarrow -1$ (blue) and $m_s=0\rightarrow +1$ (red) transitions are pumped using microwaves. Solids lines are fits of the form $A+B\sin(4\alpha+C)$.}
    \label{fig3}
\end{figure}
 

Figure \ref{fig3} shows the peak-to-peak amplitude of FDMR as a function of the angle ($\alpha$) between the fast axis of a quarter waveplate ($532~$nm) and the axis of the linear polarizer for optical transitions from the $m_s=\pm1$ states in the ground state manifold. In accord with the optical selection rules, as the light becomes more and more $\sigma_+$ polarized, the strength of the FDMR due to $m_s=-1$ reaches its peak while that due to $m_s=+1$ reaches its minimum. The situation changes when light becomes $\sigma_-$ polarized and the FDMR due to $m_s=+1$ reaches its maximum while that due to $m_s=-1$ is minimized. This demonstrates the existence of spin-orbit coupling at room temperature. We get a contrast of $60\%$ between the maximum and minimum strength of the FDMR peaks instead of the ideal $100\%$. This might be due to the ensemble nature of our experiment and the experimental imperfections including light not being completely circularly polarized. Experimental imperfections might also explain the slight offset ($12^o$) in reaching the minimum (maximum) strength of FDMR peaks (Fig. \ref{fig3}) at $57^o$ instead of $45^o$. The broader question, however, is which of the possible transitions between the ground and excited states are responsible for the behavior we see in Fig. \ref{fig3}. The selection rules \cite{Maze_2011} indicate that when light is $\sigma_+$ polarized, $m_s=-1\rightarrow \{A_1,~A_2\}$ and $m_s=+1\rightarrow \{E_1,~E_2\}$ transitions are allowed implying FDMR maxima for both $m_s=\pm1$. Experimentally, we find that only FDMR associated with $m_s=-1$ is maximized. It might be that the selection rules are not completely followed at room temperature, and the oscillator strengths of these transitions are unequal. The latter point might also explain the often weak and fluctuating nature of the FDMRs owing to the $m_s=+1$ state that we measure. It might also be the case that due to thermal averaging at room temperature \cite{Rogers_2009}, not all transitions demonstrated at low temperature \cite{ToganE2010} are affected equally and only the $m_s=\pm\rightarrow A_2$ transition survives, which is the highest energy level in the $^3E$ manifold and robust against strain \cite{Maze_2011,ToganE2010,Doherty2013}. The $m_s=\pm 1\rightarrow A_2$ transition is then responsible for the data shown in Fig. \ref{fig3}.  

\begin{figure}
    \centering
     \includegraphics[width=8.5cm]{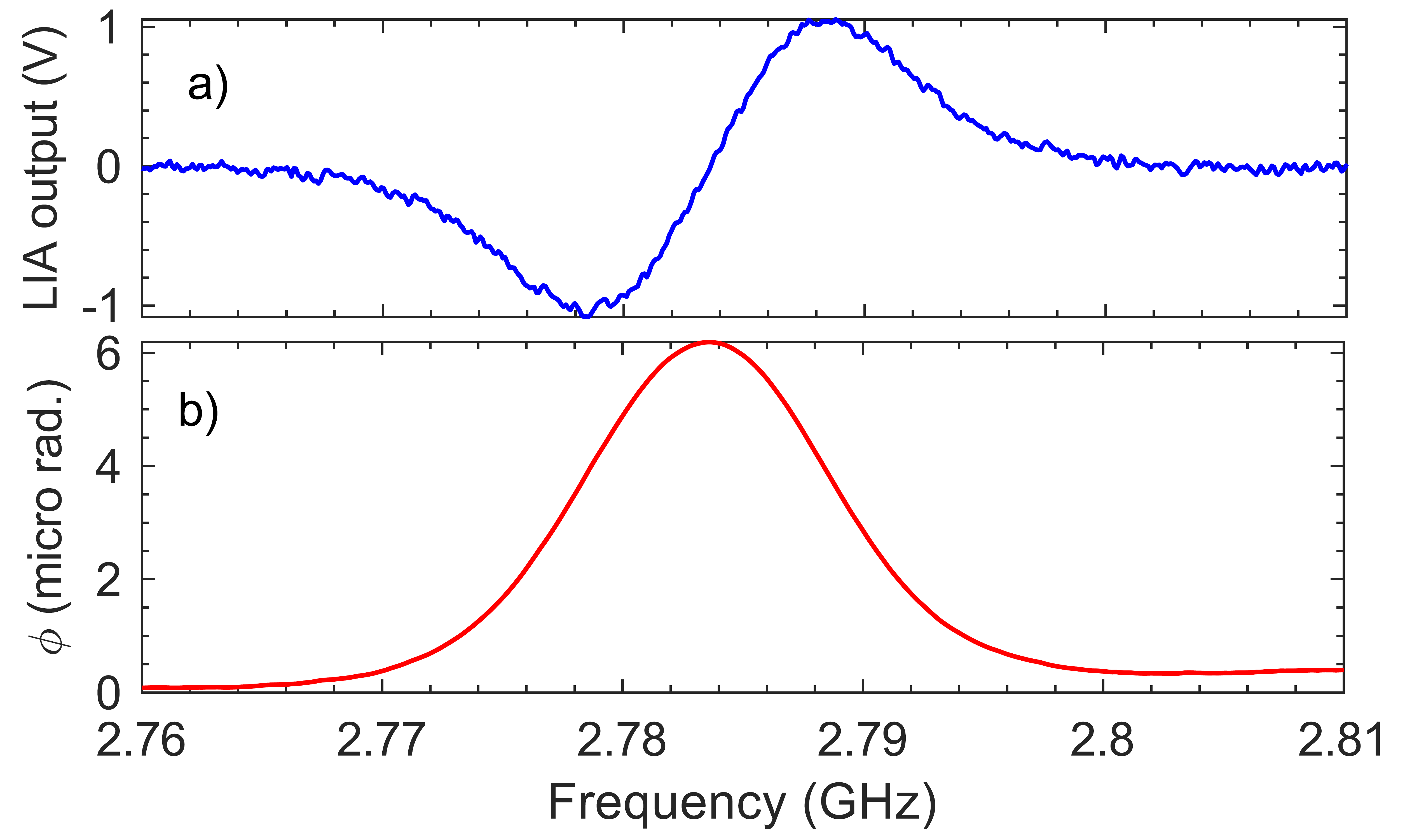}
    \caption{a) Lock-in apmplifier output as a function of the microwave frequency when the $m_s=0\rightarrow -1$ transition is pumped. b) The plane of polarization rotation angle corresponding to the FDMR signal shown in a (see text for more information).}
    \label{fig4}
\end{figure}

The rotation angle $\phi$ of the plane of polarization of the linearly polarised green light after it has travelled through the diamond can be found from the balanced detector signal \cite{BudkerRevModPhys2002} $\phi \approx \frac{\Delta P}{2P}$, where $P$ is the input power to the polarimeter, $\Delta P$ is the difference signal of the balanced detector and $|\phi|\ll \pi/4$ is assumed. The input power to the detector can be directly measured. However, due to the smallness of $\phi$, $\Delta P $ is small and often perturbed by the laser noise and drift. To increase the signal-to-noise ratio, we use a lock-in amplifier. The output of the lock-in amplifier is appropriately calibrated and subsequently integrated to find $\phi$. Figure \ref{fig4}a shows the raw output of the lock-in amplifier while panel b shows $\phi$ as a function of the microwave frequency when the $m_s=0\rightarrow -1$ transition belonging to the $[111]$ is pumped. On microwave resonance, where the highest number of NV centers are in the $m_s=-1$ state, the plane of polarization of the green laser is rotated by $6~$micro radians. When the $m_s=0\rightarrow +1$ transition is pumped, the plane of polarization of the green laser is rotated by the same amount but in the opposite direction.

The plane of polarization rotation angle $\phi$ can be calculated from the perspective of circular birefringence \cite{BudkerRevModPhys2002,ShahPRA2009}. In particular, when the $m_s=0\rightarrow -1 (+1)$ transition is pumped using microwaves, the NV centers become spin polarized i.e., $\langle S_z\rangle < 0 ~(>0)$, and the left and right circularly polarized light encounters different indices of refraction. This difference in index of refraction rotates the plane of polarization of the linearly polarized light by 
 \begin{eqnarray}
 \phi = 2n_t k L\rho_n \Re{\zeta}\langle S_z\rangle\hat{z}\text{\Large $\cdot$}~ \hat{k},
 \end{eqnarray}

 where $n_t$ is the number of times probe light of wavelength $\lambda$ travels through the diamond of length $L$, $k=2\pi/\lambda$, $\rho_n$ is the number density of NV centers, $\zeta=\frac{r_ef_sc^2L(\nu)}{4\pi \nu}$ with $c$ being the speed of light and $r_e$ being the classical electron radius. $f_s$ and $L(\nu)=(\nu-\nu_0)/(\nu-\nu_0+i\Gamma)$ are the oscillator strength and the absorption spectral profile of an NV center with $\nu=c/\lambda$ being the laser frequency, $\nu_0$ is the optical resonance frequency of the NV center and $\Gamma$ is the linewidth of the optical resonance. $\langle S_z\rangle=\sum_sm_sP_s$ is the average spin polarization of NV centers along the $z~$axis with $P_s$ being the probability of an NV center being in the $s$ state. With the theoretical and experimental values of $\phi$ at hand, the number of NV centers contributing to the Faraday effect can be estimated. The absorption profile of NV centers is broad and is not well defined at room temperature. For our estimation we consider the NV center to have a Lorentzian absorption profile with $\Gamma \approx 10^{14}~$Hz ($\approx 98~$nm) and $\nu_0=5.4\times 10^{14}~$Hz ($555~$nm) \cite{Davies1972A,Fraczek:17}, $f_s=0.02$  \cite{SantoriPRL2006}. From the rate-equations-based simulation \cite{MansonPRB2006}, for our microwave and laser power, we find $\approx 36~(1)\%$ of the NV population is in the $m_s=-1 (+1)$ state. The net spin polarization contributing to the Faraday effect is $\langle S_z\rangle=-0.35$. With $\phi= 6\times 10^{-6}~$radians. at resonance (Fig. \ref{fig4}b) and taking into account that we are only exciting $\approx 1\%$ of the NV centers due to the limited amount of green laser power available (see supplementary information), we get $\rho_n=1.1\times 10^{22}~$m$^{-3}$. From our EPR measurements, the density of NV centers is $(4.01\pm 0.36)\times 10^{22}~$m$^{-3}$. Considering the uncertainty in different parameters, including the NV absorption profile and oscillator strength, $\rho_n$ found from $\phi$ is reasonable. Using this $\rho_n$ and taking the laser illuminated volume of the diamond as that of a cylinder of diameter $1~$mm and length $L=2~$mm, we calculate the number of NV centers interacted with the green probe beam is $n_s=2.1\times 10^{10}$. In other words, a single photon has interacted with $250$ centers while traveling through the diamond.

\begin{figure}
    \centering
    \includegraphics[width=8.5cm]{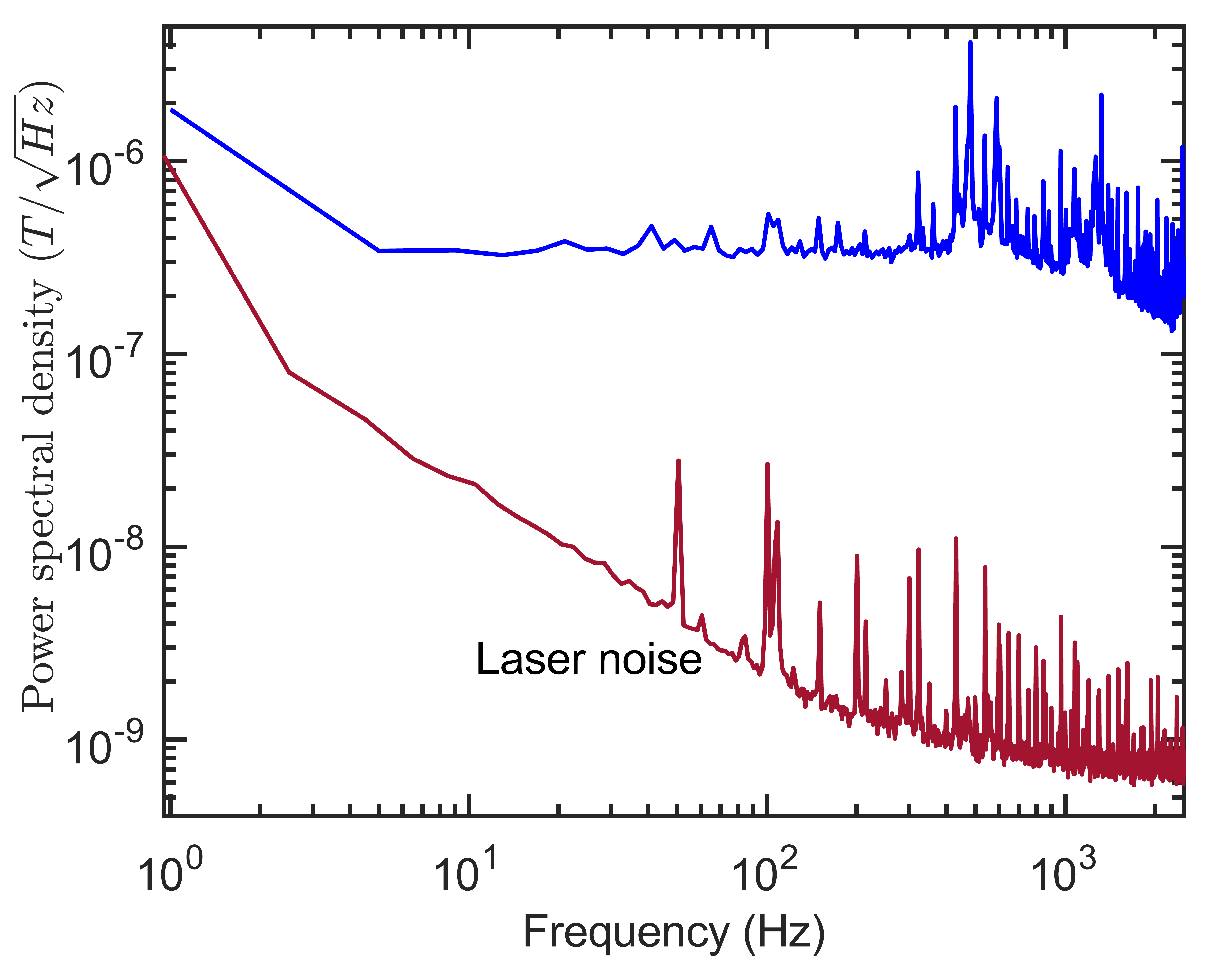}
    \caption{Sensitivity of the Faraday effect-based magnetometer. The blue line shows the Fourier transform of the lock-in amplifier output after appropriate calibration.}
    \label{fig5}
\end{figure}

The effect of the spins on the probe light's plane of polarization can be used for magnetometry \cite{KominisNat2003,BudkerRevModPhys2002} and to perform quantum nondemolition measurements \cite{ShahPRL2010,TakeiPRA2010,TakahashiPRA1999,AtatureNatPhys2007,BuckleySci2010}. Here, we develop a magnetometer. To perform magnetometry, we use the $m_s=0\rightarrow -1$ transition and continuously drive it using resonant microwaves. The green laser power after the polarizer was $1.1~$W while the microwave power was set to $33~$dBm at the input of the antenna. Magnetic fields to be detected shift the resonance frequency of the $m_s=0~\rightarrow~m_s=-1$ transition via the Zeeman effect. This changes the NV center population in the $m_s=-1$ state. This, in turn, changes $\phi$, which then perturbs the balance of the balanced detector. The lock-in amplifier, therefore, provides a modified voltage signal. With an appropriate calibration, the unknown magnetic field can be found from the lock-in amplifier (LIA) signal. Figure. \ref{fig5} shows the Fourier transforms of the calibrated outputs of the lock-in amplifier. Between $10$ and $200~$Hz, our magnetometer has a sensitivity of $350~$nT$/\sqrt{Hz}$. The spin-projection-limited minimum detectable magnetic field \cite{KominisNat2003,AcostaAPL2010,BudkerRevModPhys2002} is 

\begin{eqnarray}
\Delta B=\frac{1}{\gamma\sqrt{n_t n_sT_2^*t}},
\label{eqn4}
\end{eqnarray}

where $n_s$ is the number of the NV center that contributed to the Faraday effect, $T_2^*$ is the intrinsic spin coherence time and $t$ is the measurement time. The intrinsic spin coherence time of our diamond is $T_2^*=21~$ns. Substituting $n_s$, $T_2^*$ and $t=0.5~$s in (\ref{eqn4}), we get $\Delta B=1.7\times 10^{-12}$T. This is significantly better than what we have measured. When we set the frequency of our microwave source away from the resonance, the noise floor remains the same, indicating that we are limited by non-magnetic noise. When we separately measure the laser noise with an appropriate calibration, the noise floor is significantly lower. The temperature of our diamond increases by $30~$K, measured using the shift in ODMR frequency \cite{AcostaRPL2010}, from room temperature at the microwave and laser power used. This is unlikely to cause a significant issue. A potential noise source is the green laser that gets trapped inside the diamond due to diamond's high refractive index \cite{BarryPRR2024}. After spending some time inside the diamond, the trapped light with arbitrary phase eventually exits the diamond, travels with the retroreflected beam, and interferes, causing a white noise-like background. By covering the diamond with anti-reflection coatings, this can be mitigated, which we aim to do in the future.

The sensitivity of our magnetometer can be improved in several ways. The laser power immediately before the PBS (Fig. \ref{fig1}b) is $120~$mW. This is significantly higher than our existing balanced detector can take. So we use a neutral density filter to attenuate the power. By using photodiodes capable of taking extra laser power, the sensitivity can be improved straightforwardly. Another major area of improvement is to use a high-quality chemical vapor deposition (CVD) diamond, which has superior NV spin properties \cite{BarryPRR2024,GrahmanPRR2023}. For example, $T_2^*=14~\mu$s has ben measured \cite{BarryPRR2024}. Additionally, the sensitivity can also be improved either by addressing more NV centers using extra green laser power or by increasing the density of the nitrogen-vacancy centers \cite{GrahmanPRR2023}. 

The Faraday effect is non-reciprocal, meaning that each time light interacts with the spins of NV centers, its plane of polarization gets rotated by another $\phi$ \cite{CrepazHerbert2015Ceam,LiRomalisPRA2011}. When we do not use a double pass geometry, the sensitivity of our magnetometer is about $\approx 1.5~\mu$T$/\sqrt{Hz}$. This implies that a substantial gain can be made by using an optical cavity \cite{CrepazHerbert2015Ceam,TakeiPRA2010}. By using a separate probe laser and a cavity around the diamond, the interaction between the probe light and the NV centers can be significantly enhanced. The cavity could be formed using two independent mirrors or by coating the diamond surfaces with mirror coating. When two mirrors are used to form the cavity, the surfaces of the diamond can be coated with an antireflection coating to enhance cavity finesse \cite{Chatzidrosos2017}. Cavities formed by directly coating the surfaces of the diamond might be more stable due to their monolithic nature. Such cavities might also be preferable for applications requiring high spatial resolutions, such as biomedical imaging, since the dimensions of such a cavity will be the size of the diamond itself. The number of times, $n_t$, probe light bounces back and forth inside the cavity and thus interacts with the NV centers is determined by the cavity finesse ($\textit{F}$) or the reflectivity $r_f$ of the cavity mirrors with $F=1/(1-r_f)$. Assuming $r_f=99.99\%$, a cavity finesse of $10^4$ is achievable \cite{PontinA.2023Scco,Delic2020}. Such a cavity will increase the sensitivity by two orders of magnitude. The maximum time for which light can be confined within the cavity to enhance sensitivity is the $\le T_2^*$ time of the nitrogen-vacancy centers.

In our current experiment, we continuously pump the NV centers using a green laser and resonant microwaves. This creates a competition between the spin polarization to the $m_s=0$ state by the green laser and the microwave excitation to the $m_s=-1$ state. This reduces the interaction time between probe light and the spin. This can be avoided by implementing pulsed operation in which a green laser pulse is used to initialize NV centers to the $m_s=0$ state and subsequently, a microwave $\pi~$pulse is applied to excite the spins to the $m_s=-1$ state. This will allow probe light to interact with the spins more efficiently and thus enhance the sensitivity. The performance of our magnetometer can also be improved by driving the hyperfine transitions \cite{GrahmanPRR2023,BarryPRR2024}. With all enhancements incorporated, NV Faraday magnetometers could achieve on the order of a femtotesla sensitivity.

In conclusion, we have demonstrated the Faraday effect at room temperature and used this effect for magnetometry. To our knowledge, this is the first time the Faraday effect has been used to develop a magnetometer using NV centers in diamond. Our measurements also show that the spin-orbit coupling in NV centers exists at room temperature, in contrast with the earlier predictions \cite{Rogers_2009}. At present, the sensitivity of our Faraday magnetometer is limited by the background caused by the trapped excitation laser, which can be mitigated using anti-reflection coatings. Using a better diamond, laser, and a cavity, sensitivities in the femotesla level seem viable. Our experiment also opens the door to quantum non-demolition measurements of NV center spin states required in some applications \cite{WanPRL2016}. Finally, our Faraday-effect-based magnetometer is amenable to different techniques such as gradiometry developed by the optically pumped magnetometer (OPM) community \cite{BudkerRevModPhys2002,Fabricant_2023}. The benefit of a solid state sensor combined with the techniques developed for OPMs can be advantageous for developing new technological platforms from biomedical imaging to magnetic navigation to fundamental physics.


\textit{Acknowledgment: We acknowledge our illuminating discussions with Prof. D. Budker. His insightful comments and suggestions have strengthen this manusript significantly. We also acknowledge our discusions with V. Acosta, P. Hemmer and K. Pauli. AR acknowledges the financial support of Warwick Innovations and an EPSRC IAA grant.}


%

\end{document}